\documentclass[useAMS]{mn2e}
\usepackage{times}
\usepackage{psfig}
\def\be{\begin{equation}}
\def\ee{\end{equation}}
\begin{document}
\title[Chromothermal oscillations.....]{Chromo-thermal oscillations and collapse of strange stars to black holes : Astrophysical Implications}

\author[Bagchi et al. ...]
{\parbox[t]{\textwidth}{
        Manjari Bagchi$^{1,2,3}$, Rachid Ouyed$^{3}$, Jan Staff$^{3,4}$, Subharthi Ray$^{5}$,
        Mira Dey$^{2,~*}$, \\ Jishnu Dey$^{2,~\dagger}$}\\
\vspace*{3pt} \\
\\ $^1$ Tata Institute of Fundamental Research, Homi Bhaba Road, Colaba, Mumbai 400 005, India.
\\ $^2$ Dept. of Physics, Presidency College, 86/1 , College Street,Kolkata 700 073, India.
\\ $^3$ Department of Physics and Astronomy, University of Calgary, Canada AB T2N 1N4.
\\ $^4$ Department of Physics, Purdue University, 525 Northwestern Avenue
West Lafayette, IN 47907-2036, USA.
\\ $^5$ Inter University Centre for Astronomy and Astrophysics, Ganeshkhind, Pune 411 007, India. 
\\ $*$ Ramanna Research Fellow, Govt. of India.
\\ $\dagger$  CSIR emeritus Scientist, Govt. of India. }
\maketitle

\begin{abstract}
{The effects of temperature on strange stars are studied and it is found  that the maximum mass of the star decreases with the increase of temperature since at high temperatures the equations of state become softer. Moreover, if the temperature of a strange star increases, keeping its baryon number fixed, its gravitational mass increases and radius decreases. This leads to a limiting temperature where it turns into a black hole. These features are due to a combined effect of the change of gluon mass and the quark distribution with temperature. We report a new kind of radial oscillations of strange
stars driven by what we call {\it chromo-thermal} instability. We also discuss the relevance of our findings in the astrophysics of core collapse supernovae and gamma ray bursts.
 }
\end{abstract}

\begin{keywords}
{compact object -- neutron star -- strange star -- black hole -- GRB -- afterglow}
\end{keywords}

\section{Introduction}
\label{sec:intro}
The concept of Strange Star (SS) is not a new one. Itoh (1970) first proposed a model for it even when the QCD theory did not reach its current status. During this period Bodmer (1971) discussed about an astrophysical object collapsed from its usual state formed by nucleonic matter.  He argued that Collapsed Nuclei have radii much smaller than ordinary nuclei. Then Witten (1984) suggested that the early universe may have undergone a first order phase transition from high
temperature dense quark phase to low temperature dilute baryonic
phase, both the phases being stable with local minima. The low
temperature phase grow and gradually occupy more than half of the
volume. At this point the high temperature regions detach from
each other into isolated bubbles. Then further expansion of the
universe results in cooling in low temperature phase. These
bubbles may survive till today in the form of strange stars. He
also conjectured that the true ground state of matter is ``
Strange Quark Matter (SQM)", not $Fe^{56}$. SQM is a bulk quark
matter phase consisting of roughly equal numbers of up, down and
strange quarks plus a small number of electrons to guarantee
charge neutrality. SQM  would then have a lower energy per baryon than
ordinary nuclei and manifests in the form of strange stars. But the SS hypothesis is not till beyond skepticism. The maximum mass-radius obtained in SS models are usually lower than that obtained from neutron star (NS) models. So SS models face difficulties in explaining the high mass stars
like PSR J0751+1807 (Nice et al., 2005) and EXO 0748-676 (Ozel, 2006) which have masses $\sim 2.1M_{\odot}$. Even only a very few stiff neutron matter Equations of State (EsoS) can explain the estimated mass-radius of EXO 0748-676 (Ozel, 2006) .
On the other hand, there are some observational features which are easier to explain with SS model rather than NS model. As examples, harmonic absorption lines in the X-ray
spectrum of compact stars like 1E 1207.4$-$5209, J1210$-$5226, RX
J1856.$-$3754 $etc$, can be explained as the consequences of
compressional modes of surface vibrations of SS (Sinha $et~al$
2003). SS can produce such oscillations since they have sharp
surfaces and are self bound with zero pressure at the surface,
where energy per baryon (E/A) is minimum and the density is
non-zero. Such harmonic compressional modes are not possible for
a neutron star because of the lack of minimum in E/A. Another observed
feature, superburst, faces difficulties in using the standard
carbon burning scenario within neutron star models. Sinha
$et~al.$ (2002) proposed an alternative scenario as the formation
of diquark pairs on the strange star surface and their subsequent
breaking - giving a continuous and prolonged emission of energy
comparable with that emitted during the superbursts.

So we are supporting the view that both SS and NS exists in nature and under suitable physical condition an NS may convert to an SS leading to energy outburst (Bombaci $et~al.$ 2004, Staff $et~al.$ 2007).
But further study is needed to pinpoint the condition which will give rise to such a transition.

With this view, we feel it is intriguing to study the properties of SS by matching
theoretical predictions (of mass, radius etc.) with astronomical
observations. One such property is the temperature dependence of
the mass and the radius of an SS as it cools (by neutrino or photon emission) or heats up (by accreting matter). In this paper, the effect of temperature on SQM EoS is studied. We show that an increase in temperature leads to a softening of the SQM EoS giving rise to a decrease in maximum mass of the SS. When the temperature drops, the EoS stiffens again. We show that this leads to {\it chromo-thermal} instability which has interesting implications in astrophysics.

There are several EoSs for SQM, such as the Bag model (Alcock
$et~al.$ 1986, Haensel $et~al.$ 1986, Kettner $et~al.$ 1995), the
perturbative quantum chromodynamics (QCD) model (Fraga $et~al.$
2001), the chiral chromodielectric model (Malheiro $et~al.$ 2003)
etc. However, in all these models, the effect of finite
temperature was not sufficiently studied. Such study has been
performed in the present work using the relativistic mean field
model (Dey $et~al$ 1998, Bagchi $et~al.$ 2006) for SQM. The other novelty in this model
is that here chiral symmetry is restored at high densities due to
the introduction of a density dependence of quark masses. The
paper is organized as follows: our model for SS is briefly
discussed in $\S$ \ref{sec:model} and $\S$ \ref{sec:sstobh} is
devoted to the study of SS properties at finite temperatures
including the {\it chromo-thermal} instability and possible
collapse of an SS to a black hole. In $\S$ \ref{sec:appl}, we
have discussed the relevance of our results in the astrophysics
of core collapse supernovae and gamma ray bursts (GRBs). We
conclude in $\S$ \ref{sec:conclude} with a suggestion on how these
oscillations can be observed.

\section{The mean field model for strange stars}
\label{sec:model}

Strange stars are more compact than neutron stars and fit into
the Bodmer$-$Witten hypothesis for the existence of SQM. In the
relativistic Hartree-Fock calculation for the SQM EoS, Dey
$et~al.$ (1998) used a phenomenological inter-quark interaction,
namely the Richardson potential (Richardson, 1979). This potential
takes care of two features of the inter-quark force,
namely asymptotic freedom (AF) and confinement with the same
scale, which is not true from theoretical considerations.

Later, SS properties have been calculated with a
modified Richardson potential having different scales for AF
and confinement. The scale values were obtained from baryon
magnetic moment calculations (Bagchi $et~al.$ 2004). In this model,
chiral symmetry restoration at high density is incorporated by
introducing density dependent quark masses. A temperature dependence of
gluon mass is considered in addition to its usual density dependence.
The gluon mass represents the medium effect, resulting in the screening
of the inter-quark interaction. The temperature effect through the Fermi
function is also incorporated and at non-zero temperature, free energy $F=E-TS$ is used instead of energy $E$ while calculating the EoS. It is also ensured that in SQM, the chemical
potentials of the quarks satisfy ${\beta}$ equilibrium and charge
neutrality conditions. The parameters of the model are adjusted
in such a way that the minimum value of $E/A$ for u,d,s quark
matter is less than that of Fe$^{56}$, so that u,d,s quark matter
can constitute stable stars. The minimum value of $E/A$ is
obtained at the star surface where the pressure is zero. However,
the minimum value of $E/A$ for u,d quark matter is greater than
that of Fe$^{56}$ so that Fe$^{56}$ remains the most stable
element in the non-strange world. With the obtained EsoS,
Tolman--Oppenheimer--Volkov equations for hydrostatic equilibrium
are solved to get the structures of the stars at different
temperatures (Bagchi $et~al.$ 2006).

\section{Temperature effect on strange star properties}
\label{sec:sstobh}

\begin{figure}
\centerline{\psfig{figure=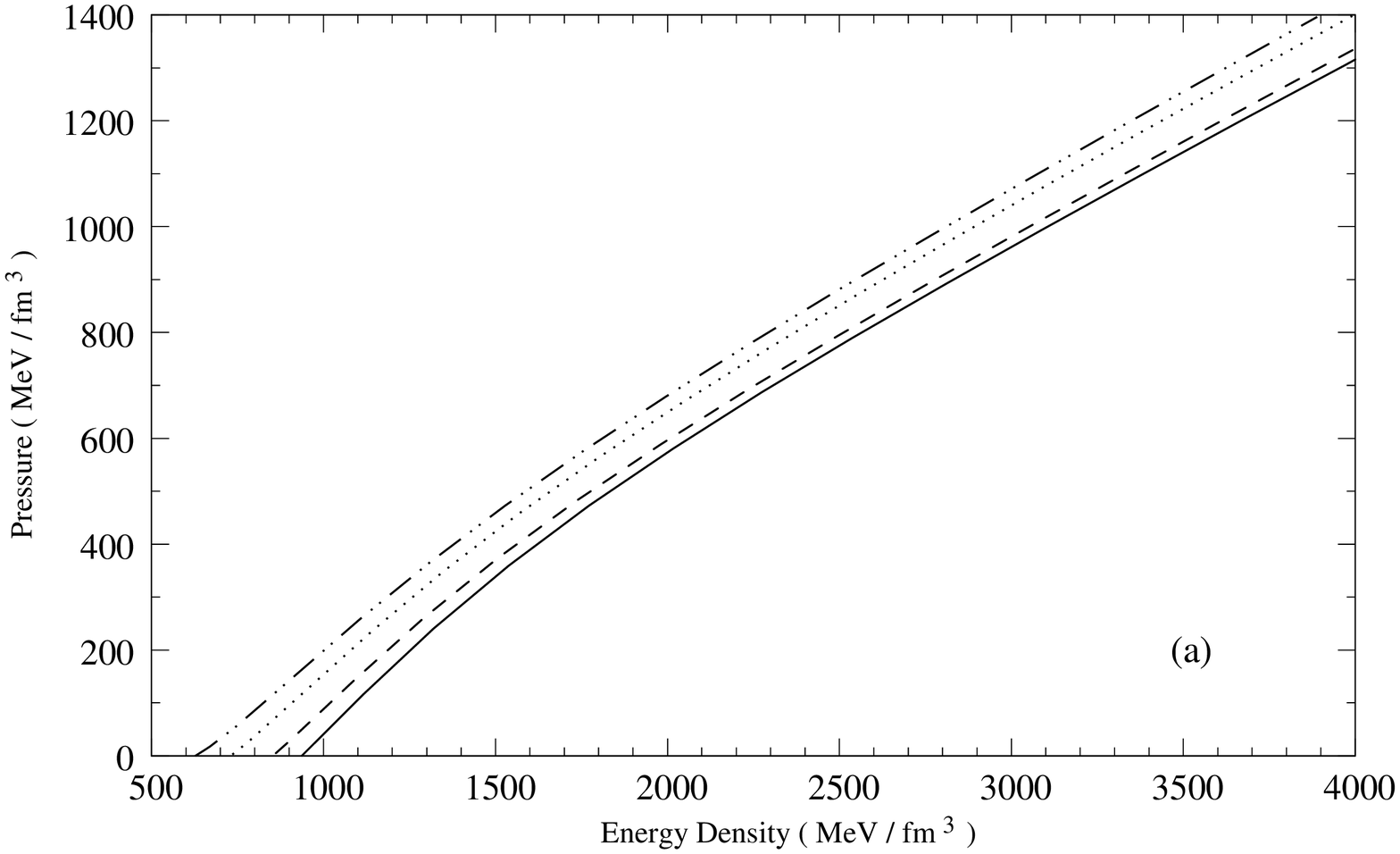,width=8cm}}
\centerline{\psfig{figure=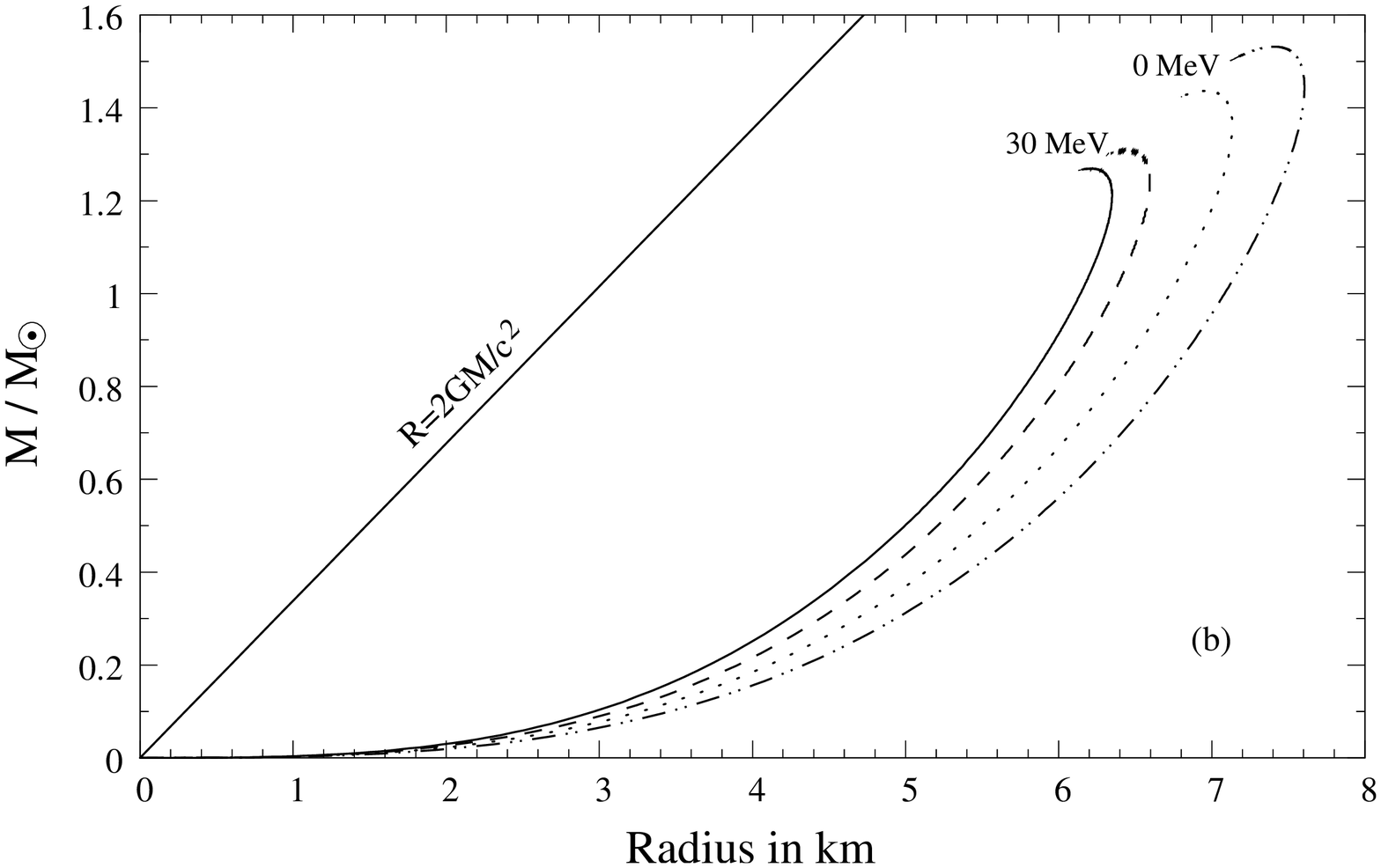,width=8cm}}
\caption{Temperature effect on strange quark matter EoSs.
Here we choose the stiffest (EoS A) and softest (EoS F) EoSs in our model.
In panel ``a'', we plot the pressure vs energy density (EoS). The upper two lines are
the EoSs at 0 MeV (dashed-dotted one is the EoS A while the dotted one is EoS F) and
the lower two lines are the EoSs at 50 MeV (the dashed one is the EoS A and the solid
one is the EoS F). Panel ``b'' shows the corresponding mass-radius curves (with the same line symbols)} \label{fig:ss}
\end{figure}

The increase of temperature softens the SQM EoS ($i.e~ p~=~p(\epsilon)$, where p is the pressure and $\epsilon$ is the energy density - see Fig.\ref{fig:ss}a) resulting in a different curve in the mass-radius plane with a lower value of maximum mass (Fig. \ref{fig:ss}b). The figures also reveal the fact that if an EoS is stiffer than another at a given temperature, then it will remain stiffer even at a different temperature. Here we have used the stiffest (EoS A) and the softest (EoS F) EoSs in our model (Bagchi $et~al.$ 2006). The change in EoS is due to a combined effect of the change of gluon mass and the quark distribution function (Fermi function) with temperature, which also changes the number
densities ($n$) of the quarks. Both the energy density and
number density increases with temperature. But the increase in
energy density is more dominant than that in number density (see Bagchi $et~al.$ 2006 for expressions of $\epsilon$ and $n$) . So energy per baryon increases with rising temperature - resulting in an increase of mass of the star for a fixed baryon number.
We find a softening of the EoS at a higher temperature, in contrast to other models showing the opposite behavior (Kettner $et~al.$ 1995). Moreover,
all models consider the effect of temperature through Fermi
function but in addition to this, our model incorporates the
temperature dependence of the interquark interaction.

In our model, while establishing beta equilibrium conditions, we have assumed that neutrinos have escaped from the star. This is true in the low temperature regime. But at high temperatures like 50-80 MeV, some of the neutrinos will be trapped inside the star which will affect the EoS. But here for simplicity, we neglect this effect as our concern is the astrophysical processes in moderately low temperature where this assumption is valid.

\begin{table}
\caption{Maximum mass strange stars at different temperatures for EoS F.}
\begin{center}
\begin{tabular}{cccc}
\hline
 Temperature  & $M_{max}$ & Radius & Baryon No.  \\
 ({\rm MeV})  & ($M_{\odot}$)  & ({\rm km})  & ($10^{57}$)  \\
\hline
 0  & 1.44 & 6.96 & 2.17  \\
 10 & 1.38 & 6.72  & 2.00 \\
20 & 1.32 & 6.44  & 1.81 \\
 30 & 1.27 & 6.21  & 1.64  \\
 \hline
\end{tabular}
\end{center}
\label{tb:maxmass}
\end{table}

\begin{figure}
\centerline{\psfig{figure=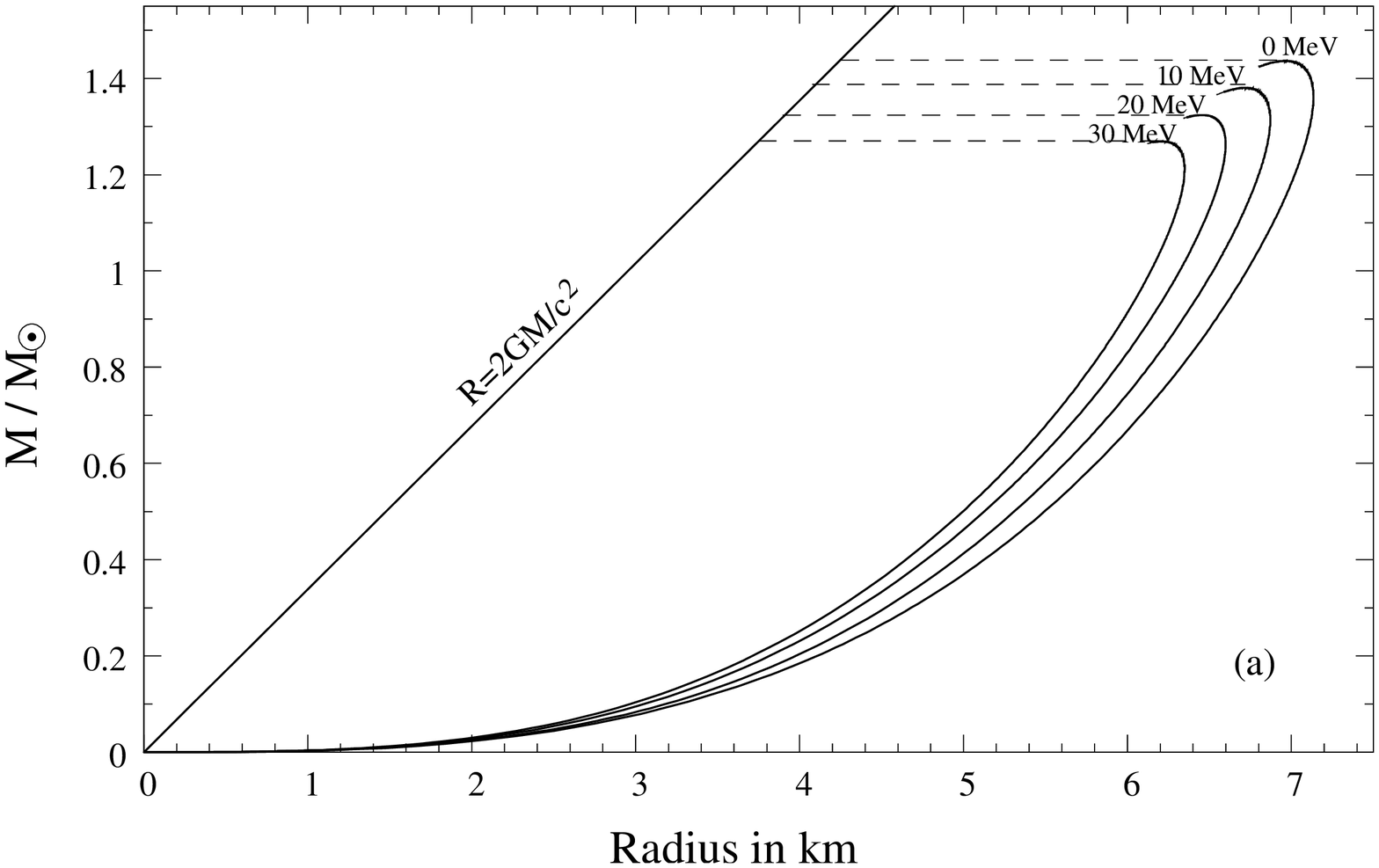,width=8cm}}
\centerline{\psfig{figure=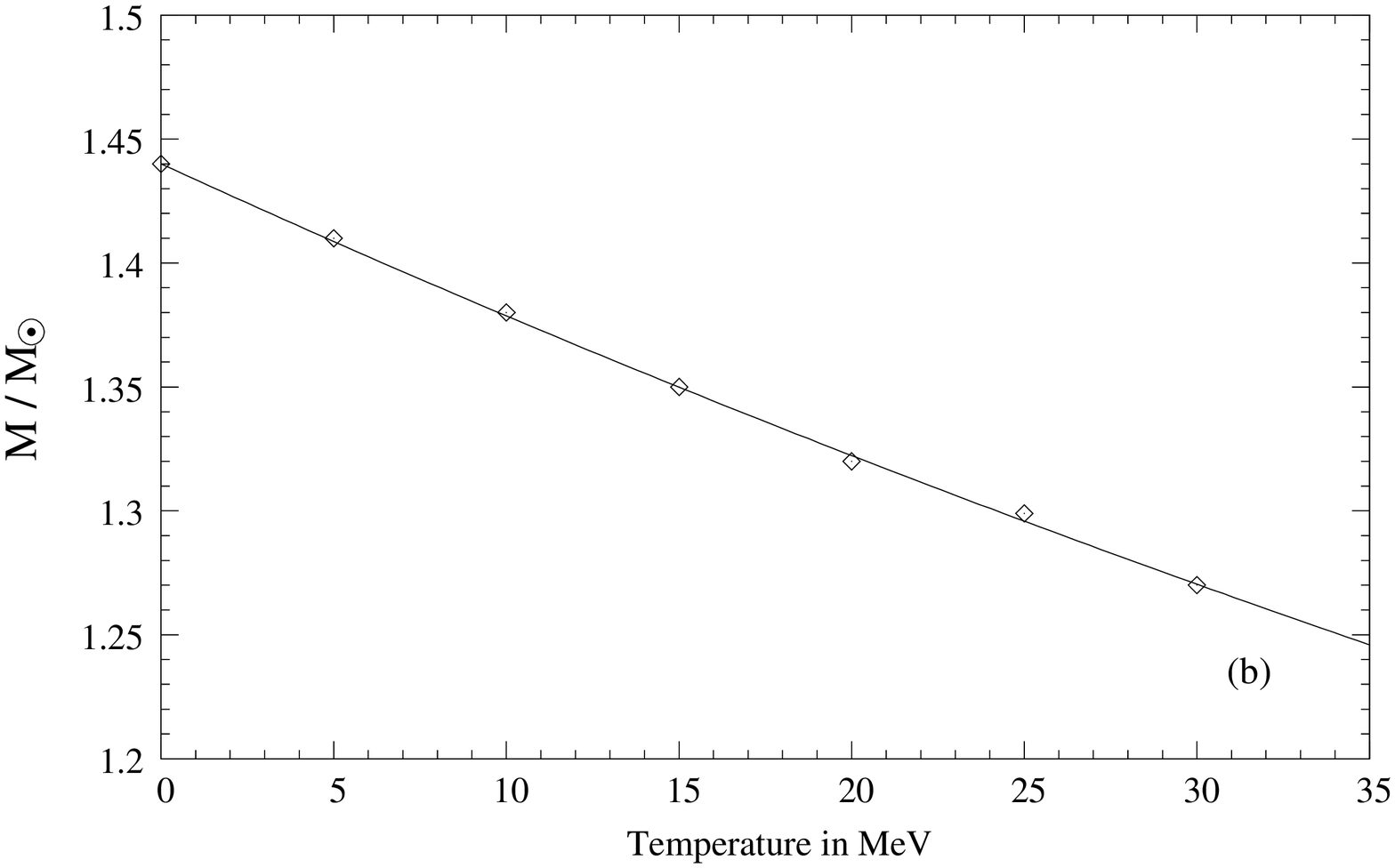,width=8cm}}
\caption{In panel ``a'', we plot strange star masses against radii at
different temperatures for EoS F. The straight inclined line on the left
represents the Schwarzchild limit so the objects located to the
left of this line are black holes. For each horizontal dashed line, the right end corresponds to the strange star and the left end corresponds to the black hole which is formed by the collapse of that
strange star. Panel ``b'' shows a fit to the maximum mass of strange stars with
temperature. Here the solid line represents the fit (eqn. \ref{eq:maxmassfit})
while the `$\diamond$'s indicate the obtained values. } \label{fig:ssbh1}
\end{figure}

\begin{figure}
\centerline{\psfig{figure=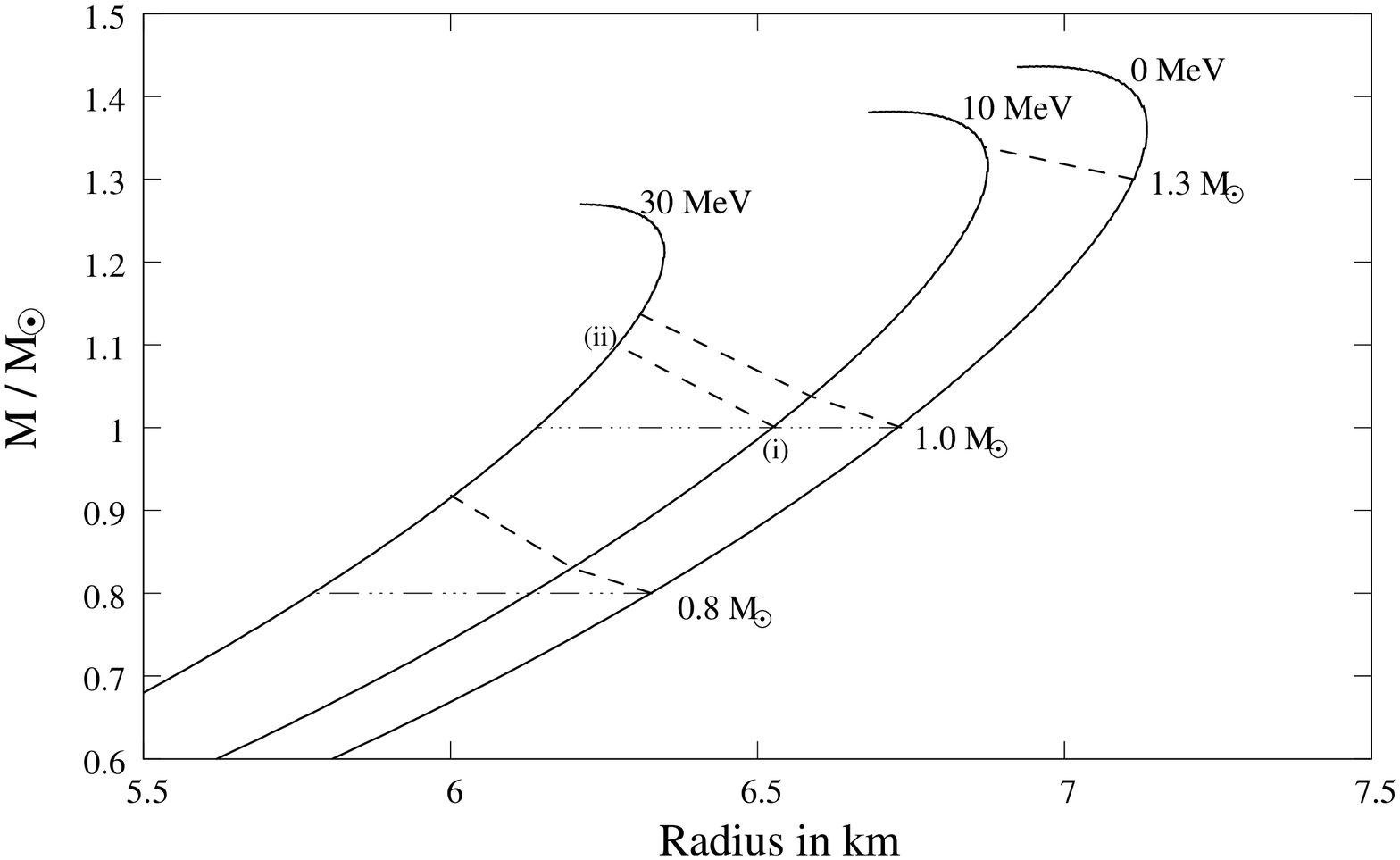,width=8cm}}

\caption{Shift of the strange stars from a low temperature to a high temperature through the dashed lines which are {\it iso-baryon} lines $i.e.$ the lines along which the total baryon number of the stars remain the same (for EoS F).} \label{fig:ssbh2}
\end{figure}

To illustrate the effect of temperature on the behavior of SS, we
choose the softest EoS (F). Fig. \ref{fig:ssbh1}a shows mass -
radius curves at different temperatures for this EoS. We also found
that for a fixed total baryon number, a higher temperature
corresponds to a higher mass of the star implying that the
effective masses of the constituent particles are greater at
higher temperatures. On the other hand, for a fixed total baryon
number, a higher temperature corresponds to a smaller radius of
the star implying that the constituent particles are closer to
each other $i.e.$ the number density is higher.

Another interesting property shown in (Fig.
\ref{fig:ssbh1}a) is the decrease of the maximum mass of strange
stars with the increase of temperature which can be fitted as :
\begin{equation}
M_{max,~T}(M_{\odot})~=~\frac{M_{max,~0}(M_{\odot})}{1+~a T_{MeV}} \label{eq:maxmassfit}
\end{equation}
where $M_{max,~T}(M_{\odot})$ is the maximum mass (in the unit of
solar mass), $T_{MeV}$ is the temperature of the star in units of
MeV and $M_{max,~0}(M_{\odot})$ is the maximum mass at a temperature of 0 MeV and $a$ is a parameter. As
an example, EOS F gives $M_{max,~0}(M_{\odot})$ $=1.4485$ and
$a~$ $=~0.0047$ \footnote{Different EoSs will give different
values for $M_{max,~0}(M_{\odot})$ and $a$. EoS A gives
$M_{max,~0}(M_{\odot})$ $=1.5317$ and $a~$ $=~0.0054$. As EoS A
is the stiffest and EoS F is the softest in our model, we
conclude that Eqn \ref{eq:maxmassfit} can be used for any EoS of
SQM with an average value of $a~$ $=~0.005$ to
find $M_{max,~T}(M_{\odot})$ at any given $T$ if
$M_{max,~0}(M_{\odot})$ is known. This may be useful for further
study of accretion onto a strange star.}. The fit quality is
good as depicted in Fig. \ref{fig:ssbh1}b where the solid
line represents the fit and the ``$+$''s indicate maximum masses
at different temperatures.

The maximum mass point in a given M-R curve also corresponds to maximum baryon number. We find that the baryon number of a maximum mass star at a certain temperature is greater than that of a maximum mass stars at higher temperatures (Table \ref{tb:maxmass}). So an SS sitting at the maximum mass point needs to decrease its baryon number in order to increase its temperature. The implausibility of this phenomenon implies that this star can only cool. If it accrets matter which generate heat ($e.g.$ Ouyed $et~al.$ 2005), then it must collapse to a black hole. However, an SS does not necessarily have the maximum mass, rather it can have any mass less than the maximum mass. Such an SS can be heated by accretion and can lead to oscillations in the star's mass and radius, what we refer to below as the chromo-thermal oscillations.

\subsection{Chromo-thermal oscillations and collapse to a black hole}

Now let us discuss the {\it chromo-thermal} instability.  Initially, the strange star was at a low temperature (the reason for this will be discussed in section 4.1) around 0-10 MeV. At such a low temperature, all the neutrinos escape from the star. Then if there is hyper-accretion onto such a low temperature SS, the star will be heated ~20-30 MeV. As this temperature is not too high, there will not be generation of neutrinos in copious amount to affect the EoS. There will be emission of gamma-ray fireball from the hot star which will halt the accretion. During this period the star cools. Then the process repeats. Here we can take the baryon number of the star to be constant for few accretion episodes as the strange quark matter is much more dense than the normal accreting matter. 

Let us first consider the case where heating and cooling are {\bf  not simultaneous} (as mentioned just above) but {\bf equal} in magnitude. In this case, as the star heats up during an accretion event,  it will move from one M-R curve at a lower temperature to another at a higher temperature following an {\it iso-baryon} line in such a way that the mass of the star increases and the radius decreases as depicted in Fig. \ref{fig:ssbh2} which shows different probable paths or {\it iso-baryon} lines (dashed lines) corresponding to different initial masses and temperatures of the star.  Then cooling starts and the star will move to lower temperature M-R curve through the same {\it iso-baryon} line and its mass will decrease and radius will increase. Since the timescales of these phenomena are set by quark interactions, the adjustment of the star's mass and radius is to a first approximation of the same order as the signal crossing time of the star or, $R/c_{\rm s}\sim 0.04$ ms, where $c_{\rm s}$ is the speed of sound in SQM (for an estimate of $c_{\rm s}$ see Sinha $et~al.$ 2004) $i.e.$ the changes in star's mass and radius are almost instantaneous. In the case of continuous accretion, these processes will repeat making the star oscillate along an {\it iso-baryon} line between two M-R curves, one at a higher temperature and the other at a lower temperature. This subsequent increase and decrease in the radius is the oscillation driven by the {\it chromo-thermal} instability discovered in the present work. After a sufficiently long time of accretion, the oscillation will move to another  {\it iso-baryon} line representing a higher baryon number.

The nature of these  oscillations is different from the well known ``radial oscillation''  arising due to perturbations in the hydrostatic equilibrium at a constant temperature (Chandrasekhar 1964). There are two fundamental differences between these oscillations and {\it chromo-thermal} oscillation. First, in the oscillations a la  ``Chandrasekhar" the mass of the star remains constant and secondly the radial oscillations are very small in magnitude. In our case, both the mass and radius oscillate and up to 10\% change in radius and mass are possible. 

For illustration purpose, let us take the softest EoS F, stiffer EsoS will give higher values of masses than quoted below. With this EoS, let us consider a strange star with initial mass $1.0~M_{\odot}$ and initial temperature 10 MeV with total baryon number $N_{B,1.0,10}=1.36677 \times 10^{57}$ (point (i) in Fig. \ref{fig:ssbh2}) as an example. When it is heated to a temperature of 30 MeV, it will come to a point  where the mass is $1.099~M_{\odot}$ (ii in Fig. \ref{fig:ssbh2}) which has the same baryon number. During this phase the star's radius would have instantaneously decreased by $\sim 4$\% and the gravitational mass increased by $\sim 10$\%. Then cooling starts and the star will return to point (i). If accretion is still occurring, the process starts again continues until the baryon number increases significantly at which point the star finds itself on another {\it iso-baryon} line. These large oscillations in radius and mass, we expect, could have unique observational signatures (see section \ref{sec:conclude}). 

This oscillation will affect observational features in different ways - the gravitational redshift $z$ increases by 22\% when a star comes from point (i) to point (ii) of the above example. Moreover, the sudden changes in the star's radius and mass  may alter the properties of the star, such as its Schwarzschild radius and its innermost stable circular orbit (ISCO), in such a way that it might affect the accretion process and lead to disc instabilities which may be observed by observing processes related to accretion.

We have also calculated the amount of energy needed for such oscillations to occur. During each half oscillation, the energy input to the star (during heating) or the energy emission from the star (during cooling) is equal to the change in its free energy ($\Delta F$). For the above example of oscillation between point (i) and point (ii), 
\begin{eqnarray}
\Delta F = F_2- F_1 = (E_2-T_2S_2)-(E_1-T_1S_1)
 \nonumber \\= (M_2-M_1) c^2 +(T_1S_1-T_2S_2) \nonumber \\= 0.03273 \times 10^{60} MeV = 0.02933~ M_{\odot}c^2  
\label{eq:free_energy}\end{eqnarray}
where $S_1$ and $S_2$ are total entropy of the star at points (i) and (ii). 
The entropy density $s(r)$ is estimated during calculation of the EsoS (Eqn. 8 of Bagchi $et~al$ 2006). s(r) in in the unit of $fm^{-3}$ and the unitless total entropy S is obtained by integrating s(r). We found $S_1=9.736\times 10^{56}$ and $S_2=29.3711\times 10^{56}$. The amount of accretion needed during heating is determined by $\eta M_{accr}=\Delta F$ where $\eta$ is the efficiency for the accreting matter to give energy to the star (as some of the energy from the accreting matter will be emitted in the form of x-rays and so on).  The exact value of $\eta$ is not known, but we are assuming that only a small portion of the energy of the accreting matter will dissipate and so taking $\eta~=0.8$, we get $M_{accr}=0.0366625~ M_{\odot}$ needed to bring the star from point (i) to point (ii). During cooling, $i.e.$ when the star comes from point (ii) to point (i),  $\Delta F =0.03273 \times 10^{60} MeV$ energy will be emitted by the energetic photons. As $0.0366625~ M_{\odot}$ normal matter contains $0.0366625 \times 1.116  \times 10^{60} /939 ~=~0.04357 \times 10^{57}$ baryons, so after two or three cycles, the star will start to oscillate along the next higher {\it iso-baryon} line. As the initial baryon number of the star was $1.36678 \times 10^{57}$ and the maximum baryon number at 30 MeV is $1.64 \times 10^{57}$, after $(1.64-1.36678)/0.04357 \sim 6$ oscillations the star will collapse to a black hole.  This is an example of an extreme case. It is more probable that the accretion rate is lower and the star is oscillating for a long time between two close temperatures say $10~MeV$ and $12 ~MeV$. In that case the change in $M$ and $R$ will be very small, difficult to detect; but the effect of this breathing process is till possible to detect by observing other properties of the star which are affected by these oscillations - like gravitational redshift, disk properties, etc.

A more complex oscillatory behavior will result in situations where  heating and cooling rates (i.e. timescales) are not equal.

The feedback mechanism between heating/cooling (the external drive) and the star's response through the chromothermal oscillations, and the consequences  for the star properties and the underlying accretion process are being currently explored in more details.

The collapse to a black hole occurs when the star reaches such a high
temperature that the maximum baryon number at that temperature is less than the baryon number of the star.  As an example, if an SS has initial mass $1.3~ M_{\odot}$ at 0 MeV (with baryon number $N_{B, 1.3, 0}~=~19.24\times 10^{56}$), it will remain an SS when heated to 10 MeV ($ N_{B, max, 10}~=20.00\times 10^{56}> N_{B, 1.3, 0}$) but will collapse to a black hole before reaching 30 MeV (as $ N_{B, max, 30}~=~ 16.40\times 10^{56}< N_{B, 1.3, 0}$).  The lower the initial mass of the strange star, the higher the temperature it can sustain. 

Throughout the work, we have assumed that the stars are isothermal even at higher temperatures.  Random motions of quarks at very high velocities (0.5-0.7) as quoted in Bagchi $et~al.$ 2007 will facilitate isothermal configurations. For theoretical interest, we will discuss isentropic configurations in section \ref{sec:isentropy}.
\section{Isentropic configurations}
\label{sec:isentropy}

\begin{figure}
\centerline{\psfig{figure=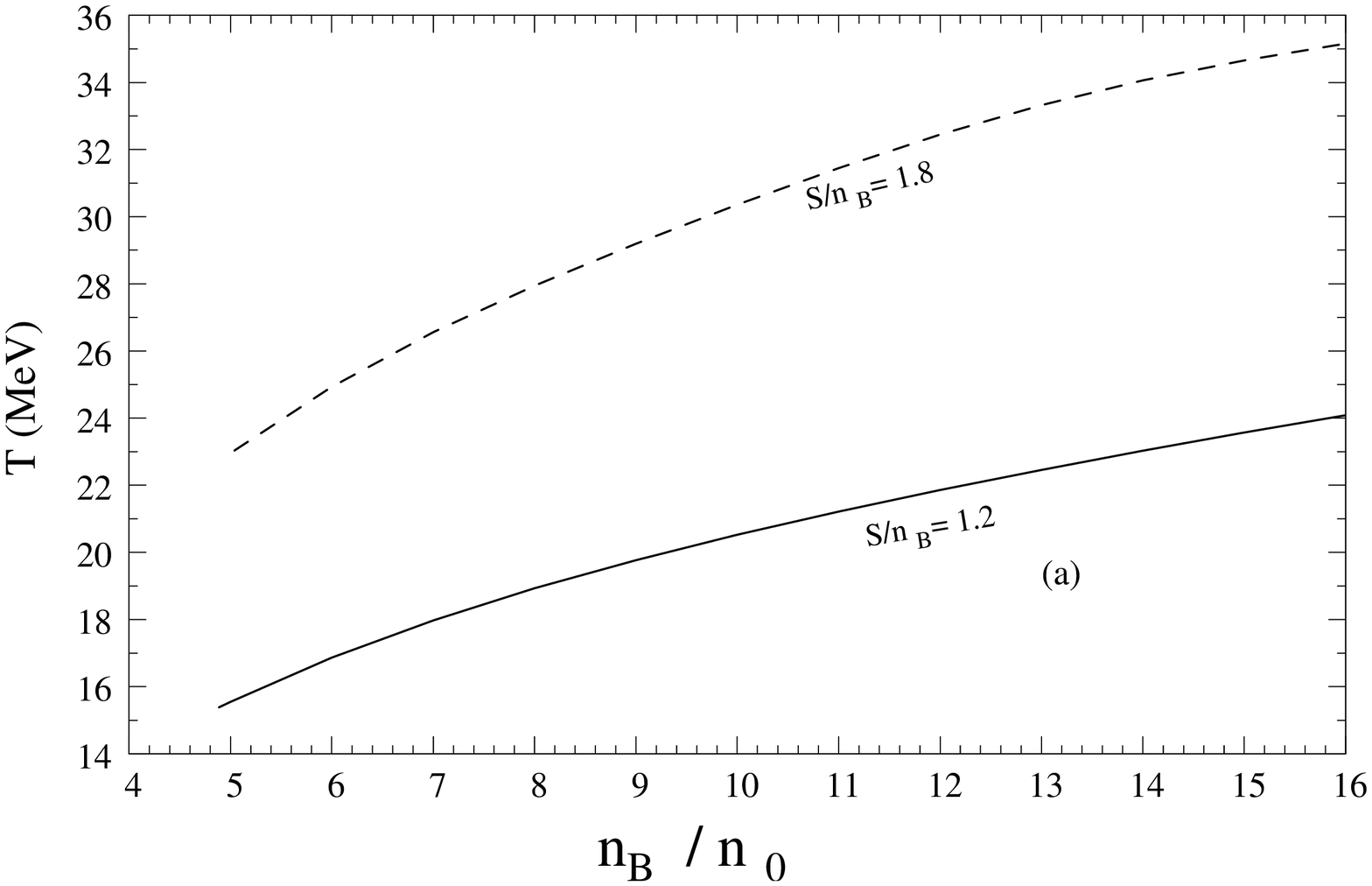,width=8cm}}
\centerline{\psfig{figure=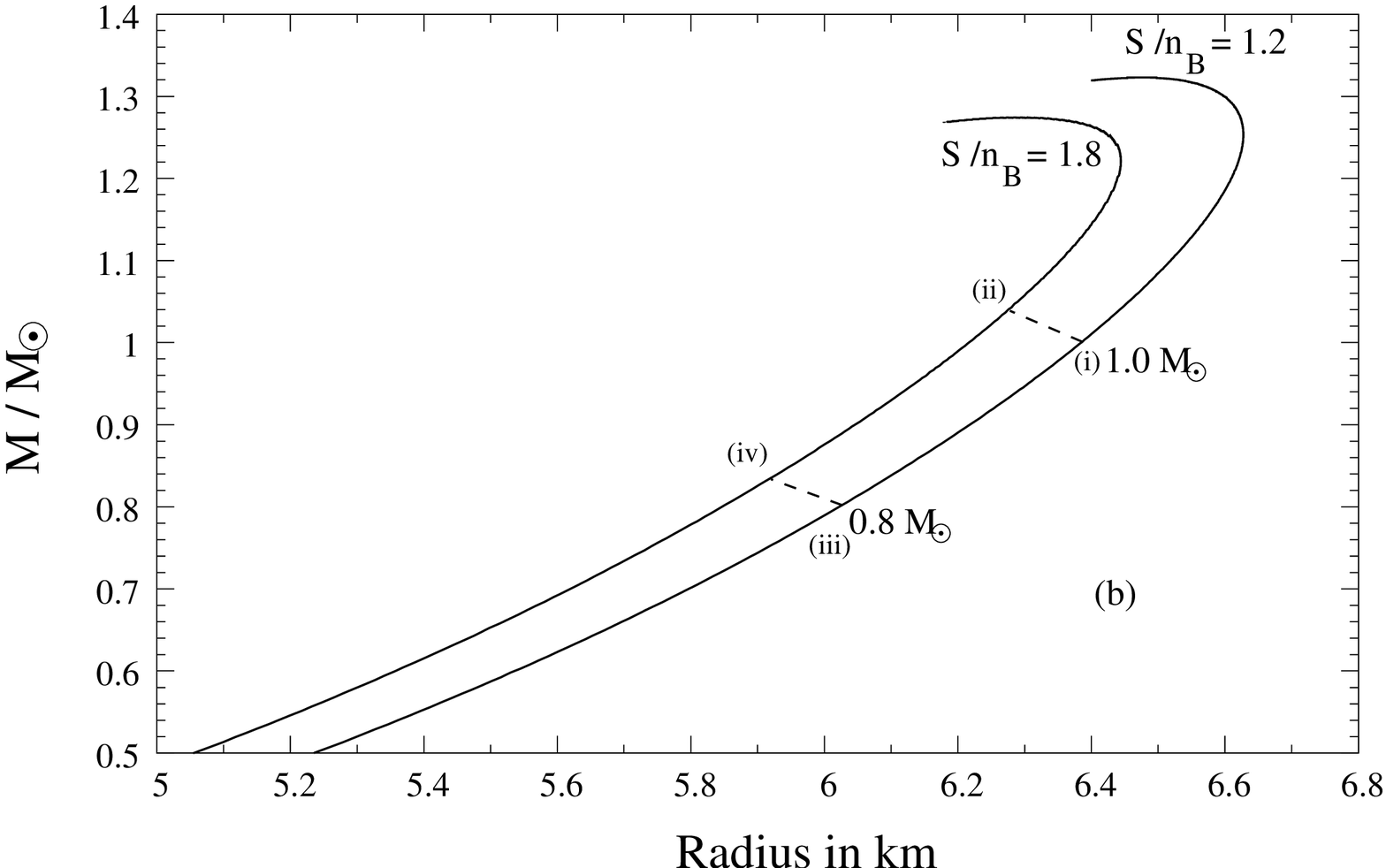,width=8cm}}
\caption{Properties of isentropic strange quark matter for EoS F in our model. In panel ``a'', we plot the temperature vs number density for $S/n_B~=$~ 1.2 and 1.8.  Panel ``b''  shows mass-radius curves for these two $S/n_B$ values and {\it chrmothermal oscillations} along {\it iso-baryon} lines (dashed lines). } \label{fig:isentropy}
\end{figure}

Till now, we have considered the strange stars to be isothermal. 
On the other hand, extensive study of neutron star evolution shows that after few seconds from its birth, a neutron star will come to an isentropic configuration (Fig. 2 of Burrows \& Lattimer, 1986). Though we have not performed such a study of strange star evolution, for the sake of completeness, here we are representing the results for isentropic configurations.

We have calculated EoSs for isentropic SQM with two chosen values of entropy per baryon ($S/n_B$) $e.g.$ 1.8 and 1.2 with the set of parameters used to get EoS F in Bagchi $et.~al$ (2006). For both the cases, temperature T increases with increasing $n_B/n_0$, and for a fixed value of $n_B/n_0$, T is higer for a higher $S/n_B$ as shown in the upper panel (panel ``a") of Fig \ref{fig:isentropy}.

The EoS becomes softer with increase of $S/n_B$ giving lower value of maximum mass.  Table \ref{tb:maxmass_isentropy} shows different parameters of maximum mass stars for these two values of 
$S/n_B$. As usual, maximum mass star contains maximum total baryon number. The lower panel (panel ``b")  of Fig \ref{fig:isentropy} shows mass-radius curves with these two values of $S/n_B$.

The figure in panel ``a" can be interpreted as the variation of temperature from surface to center as we found that for both values of $S/n_B$, the value of $n_B/n_0$ at the surface is $\sim5$ and at the center it is $\sim14$ for a maximum mass star. So we can say, for a maximum mass star, T varies from 34 Mev to 23 Mev from center to surface for $S/n_B= 1.8$ and T varies from  23 MeV to 15 Mev from center to surface for $S/n_B= 1.2$. These values are not too different from what Pons $et.~al$ obtained for proto-neutron stars (Fig 3 of Pons $et.~al$, 1999).

It is clear from table \ref{tb:maxmass_isentropy} that a maximum mass strange star can collapse to a black hole if it is heated by accretion $i.e.~ S/n_B$ increases as the maximum baryon number allowed for a higher $S/n_B$ is less than the baryon number for a maximum mass star at a lower value of $S/n_B$. 

Chromothermal oscillations along any {\it iso-baryon} line can occur here also if initially a strange star has its mass less than the maximum mass.  The figure in panel ``b" shows two such examples, oscillation can occur between points (i)-(ii) and (iii)-(iv). When a star comes from point (i) having mass $1.0~ M_{\odot}$ and radius 6.386 km to point (ii) having mass $1.04~ M_{\odot}$ and radius 6.274 km by changing $S/n_B$ value from 1.2 to 1.8 and keeping its total baryon number fixed at $1.318 \times 10^{57}$, its mass increases by an amount of $3.88\%$ and radius decreases by an amount of $1.82\%$, similarly when the star comes from point (iii) having mass $0.802~ M_{\odot}$ and radius 6.036 km to point (iv) having mass $0.854~ M_{\odot}$ and radius 5.96 km by changing $S/n_B$ value from 1.2 to 1.8 and keeping its total baryon number fixed at $1.026 \times 10^{57}$, then its mass increases by an amount of $6.46\%$ and radius decreases by an amount of $1.26\%$.
The change in free energy (see eqn \ref{eq:free_energy}) is $\Delta F~=~0.044~{\rm MeV}~=~0.0395~M_{\odot}c^2$ for a transition (i) - (ii), and it is $\Delta F~=~0.0399~{\rm MeV}~=~0.0357~M_{\odot}c^2$  for a transition (iii) - (iv). These values are equal to the amount of accretion needed when the star is being heated and comes from a lower $S/n_B$ curve to a higher $S/n_B$ curve i.e. transition from point (i) to (ii) or (iii) to (iv). Alternatively, these are the amount of energy emitted by the stars during cooling i.e. transition from a higher $S/n_B$ curve to a lower $S/n_B$ curve - like from point (ii) to (i) or from point (iv) to (iii).

\begin{table}
\caption{Maximum mass strange stars at different $S/N_B$ for EoS F.}
\begin{center}
\begin{tabular}{cccccc}
\hline
 $S/n_B$  & $M_{max}$ & Radius & $T_{center}$ &$T_{surface}$ & Baryon No.  \\
   & ($M_{\odot}$)  & ({\rm km})  &  ({\rm MeV}) &  ({\rm MeV}) & ($10^{57}$)  \\
\hline
 1.2  & 1.32 & 6.47 &34.06 &23.04 & 1.83  \\
 1.8 & 1.27 & 6.28  &23.23 & 15.38 & 1.68 \\
 \hline
\end{tabular}
\end{center}
\label{tb:maxmass_isentropy}
\end{table}

\section{Astrophysical Implications}
\label{sec:appl}
\subsection{Core collapse supernova}
In a core collapse supernova, people guess the possibility of
production of either a neutron star or a strange star or a black
hole. We model that first a neutron star is born during the supernova, which rapidly comes to a low temperature by neutrino cooling and then it converts to a strange star (e.g. Staff $et~al.$ 2006). The intermediate
neutron star stage may lead to a delay between a supernova and
the  subsequent strange star formation which has interesting
implications to the model of GRBs involving strange stars (Staff
$et~al.$ 2007).

Here we should remember that our EsoS for SQM, which ignore neutrino effects,  might not be appropriate at a very high temperature created by the supernova. But as we are modeling that first a neutron star is born during the supernova and then it converts to a strange star at a low temperature, the simplification in the EoS does not affect the physics.

\subsection{Gamma ray bursts}

The recent launch of Swift satellite has revealed some new
interesting features of the GRB afterglows which can not be
easily explained by current models. If a strange star form inside a
collapsar then a new model involving a three step process in the
framework of the ``Quark Nova'' scenario (Ouyed $et~al.$ 2002;
Ker\"anen $et~al.$ 2005) is one possible way of explaining recent
swift data. In this model, the conversion of a neutron star to a
strange star followed later by the  conversion of the  strange
star to a black hole -- as suggested in this paper -- is
involved. Here GRB occurs due to the jet from the accreting strange
star (Ouyed $et~al.$ 2005). Depending on the initial mass, temperature and the accretion rate, a strange star may collapse to a black hole and
the accretion onto the black hole gives rise to an ultra-relativistic jet causing the giant flares observed in some GRB afterglows (see Staff  $et~al.$ 2007 for details).

\section{conclusion}
\label{sec:conclude}

In this paper we studied the effects of temperature on the EoS of
SQM. A novelty in our model is the increase of the softness of the EoS at high temperatures. This feature reveals the dependence of the 
maximum allowable mass on the temperature and the change of the star's mass and radius with temperature. This suggests that the collapse of an SS
to black hole is more intricate than previously thought and might have interesting astrophysical applications specifically in situations involving accreting SS with applications to GRBs.

We have also isolated a new type of radial oscillations that we
termed  {\it chromo-thermal}. As the increase in mass is accompanied
by the decrease in radius and vice versa, the value of
gravitational redshift $z$ will also oscillate. This effect could
in principle be detected through spectral or timing
analysis techniques and should put our model to test.

\section*{Acknowledgments}
The authors thank P. Jaikumar, W. Dobler, N. Wityk and B. Niebergal for useful comments. \\ \\
This research was partly supported by the Natural Science
and Engineering Research Council of Canada (NSERC) and the Department of
Science and Technology (DST), Government of India.


\end{document}